\documentclass[pre,amsmath,twocolumn, amssymb,notitlepage,floatfix]{revtex4}

\usepackage{dcolumn}
\usepackage{lipsum}
\usepackage{graphicx}
\usepackage{placeins}
\usepackage{float}
\usepackage{bm}
\usepackage[T1]{fontenc}
\usepackage{color}
\usepackage{url}
\usepackage[bf]{subfigure}
\usepackage{rotating}
\usepackage{amsfonts}
\usepackage{multirow}	
\usepackage{xcolor}
\usepackage[utf8]{inputenc}

\begin{document}

\title{Distance-Based Network Partitioning}

\author{Paulo J. P. de Souza$^{1}$\footnote{Electronic address: paulo.paulino.souza96@gmail.com}}
\author{Cesar H. Comin$^{2}$}
\author{Luciano da F. Costa$^1$}

\affiliation{$^1$Instituto de Física de São Carlos, Universidade de São Paulo, São Carlos, São Paulo, Brazil\\
$^2$Universidade Federal de São Carlos, São Carlos, São Paulo, Brazil}
\begin{abstract}
A new method for identifying communities in networks is proposed. Reference nodes, either selected using a priory information about the network or according to relevant node measurements, are obtained so as to indicate putative communities. Distance vectors between each network node and the reference nodes are then used for defining a coordinate system representing the community structure of the network at many different scales. For modular networks, the distribution of nodes in this space often results in a well-separated clustered structure, with each cluster corresponding to a community.  One interesting feature of the reported methodology for community finding is that the coordinate system defined by the seeds allows an intuitive and direct interpretation of the situation of each node with respect to the considered communities.  The potential of the method is illustrated with respect to a community detection benchmark, a spatial network model and to city streets networks.

\begin{description}
\item[Keywords] Complex networks, network partitioning, coordinate system.  
\end{description}

\end{abstract}

\maketitle

\section{Introduction}

One of the interesting properties of several real-world complex networks --- such as scientific collaborations, brain networks, social and economical networks --- concerns their modular structure~\cite{fortunato2010community}.  Modularity is important from both topological and dynamical points of view. Topologically, communities correspond to the partitioning of the network into major groups of reference, with potential for revealing much about the possible origin of the communities as well as the behavior of different dynamics in the network.  Indeed, modularity influences dynamics, because it tends to constrain dissemination of activation inside each module (e.g.~\cite{costa2009beyond}).  It was also shown that the all important issue regarding the interplay between topology and dynamics is strongly influenced by modularity, in the sense that different types of relationships can be observed within communities of a \emph{same} network~\cite{comin2014random}.  

For all its importance and promises, modularity remains a challenge as a consequence of the difficulty of, given a network, to identify its respective communities.  Indeed, many are the reported approaches proposing new methods of community detection (e.g.~\cite{fortunato2010community, lancichinetti2009community, fortunato2007resolution}).  Part of the difficulty in finding communities can be better understood by taking into account the direct analogy between this task and the problem of \emph{clustering}, or \emph{unsupervised classification}, in the research area known as pattern recognition (e.g.~\cite{nasrabadi2007pattern, tou1974pattern, bishop2006pattern}).  The main problem in clustering concerns the diversity of manners in which a cluster can appear or be defined.  Basically, a cluster is a subset of the objects so that its elements are more similar (closer) one another than with objects in the remainder of the set.  The multiplicity of ways to define a cluster can start to be appreciated by observing that there is an infinite number of possible distance types in a geometric space (e.g. Euclidean, Mahalanobis, cityblock, etc.).  Specific metrics will possibly be more compatible with specific types of data.  

Another problem in defining clustering is the criteria to be adopted for deciding if each of the given objects belong or not to each cluster.  This issue can be effectively treated by mapping the objects into a density space, so that clusters become associated to density extremes. However, as a consequence of real-world complexity, these extremes will appear along a whole set of spatial scales.  These two main types of problems, choice of metrics and cluster identification, are directly inherited by community detection approaches.  

A third important aspect regarding community detection is related to the \emph{conceptual interpretation} of the adopted definition of modularity.   For instance, spectral methods of community detection are based on eigenvalues and respective eigenvectors that are not always directly related to topological properties underlying modularity.

The current work addresses the problem of community detection through an analogy with prototype-based supervised classification.  More specifically, a prototype node is assigned to each community, and used as a reference from which respective distances are calculated. Our motivation to define a coordination system to each node resembles to graph embedding  theory that are used by several methods to network partitioning like node2vec~\cite{grover2016node2vec} and SDNE~\cite{wang2016structuralSDNE}. The proposed method tends to be simple and its coordinate system has a physical and intuitive interpretation that highlights its potential. 

In this paper we present several interesting applications of the proposed method, starting with the corrected partitioning of the Zachary's karate club network, and then proceeding to a systematic comparison between this method and some traditional approaches for network partition using a geographical benchmark.  The obtained modularity resulted to be comparable, and sometimes better, than other methods in the literature.  In addition, the proposed methodology is also illustrated for city's network conurbation.

\section{Basic concepts}

\subsection{Modularity}\label{sec:mod}
Modularity quantifies how the connections of a network concentrate within subsets of nodes known as modules or communities~\cite{newman2006modularity}. The modularity index adopted in this work has maximum value of one.  Near zero values are obtained for loose partitions. The adopted modularity is given by Eq.~\ref{eq:modularity}

\begin{equation}\label{eq:modularity}
Q = \frac{1}{2M}\sum\limits_{ij}\left(a_{ij} - \frac{k_i k_j}{2M}\right)\delta_{I_i,I_j} \, ,
\end{equation}
where $M$ is the total number of edges in the network, $\delta_{I_i, I_j}$ is the Dirac function,  $a_{ij}$ is the element of row $i$ and column $j$, and the term $\frac{k_iK_j}{2M}$ is the expected number of connections between nodes with degree $k_i$ and $k_j$ known in the literature as the NG null model~\cite{newman2006modularity}.

\subsection{Adopted Community Detection Methods}
In order to validate the proposed method, we performed comparisons with some traditional community detection methods.  The walktrap algorithm~\cite{pons2005computingWalktrap} is based on a random walk taking place on a given network.  The method's basic idea is that the transition probabilities of the random walk tend to be similar for nodes in a same community. The algorithm successively merges nodes in the network having similar transition probability vectors, defining many different partitions of the network. It then chooses the partition that maximizes the modularity. The other community method adopted in this work is the Infomap~\cite{rosvall2007mapsInfomap}. The method tries to compress the information generated by a dynamics taking place on the network, such as a random walk. This is done by associating unique code words to each community, while reusing code words for nodes inside different communities. The third method we use, called label propagation~\cite{raghavan2007nearLabelProp}, is based on a label exchange dynamics taking place in the network. Initially, a set of labels is assigned to the nodes in the network. Then, each node changes its label to the label occurring with the highest frequency among its neighbors. The process is repeated until a convergence criterion is met. Labels can disappear during the evolution of the dynamics, leading to a new set of labels that provides a better representation of the communities in the network. Prior information about the communities can be given as an input to the algorithm. For instance, unique labels can be associated to seed nodes, while the remaining nodes are initially unlabeled. 

\section{The Distance-Vector Method}

The suggested method requires the definition of references nodes. These central nodes can be defined using a priory information about the network or according to measurements such as the degree, clustering coefficient, accessibility~\cite{costa2007characterization, viana2011many}, and/or geographic characteristics of the network.  The former situation can often be of practical interest, such as when prior information about the modularity has to be taken into account. For example, in geographic networks it is expected that the seeds coincide with the geographic center of a community. Also, in images networks~\cite{chalumeau2009complexIM2NET} applied in segmentation, that the seeds should coincide with the regions centers. 

Given the reference nodes, a distance vector $D_i$ is defined for each node $i$ in the network. The $j$-the element of $D_i$ contains all the distances between node $i$ and the reference node $c_j$ times a weight factor $\alpha_{c_j}$. That is

\begin{equation}
\vec{D}_i = (\alpha_{c_1}d_{ic_1}, \alpha_{c_2}d_{ic_2},...,\alpha_{c_M}d_{ic_M}),
\end{equation}
where $M$ is the number of reference nodes. Distance $d_{ic_j}$ can be topological or spatial. 

Each of the reference nodes can be understood as a kind of coordinate axis in a multidimensional space  $\Omega$.  The consideration of the several distances, taking into account their multiplicity, provides robust indication about the position of each of the network nodes with respect to the overall network structure, typically inducing respective well-separated clusters in $\Omega$.  Each of these clusters tend to correspond to one of the original communities.  The assignment of each node to each of the communities is done by taking into account the minimum topological or spatial distance with each of the reference nodes. Ties were decided by comparing the sum of distances of the neighbors of the node to each center.
\begin{figure}
	\begin{center}
	\includegraphics[scale=0.5]{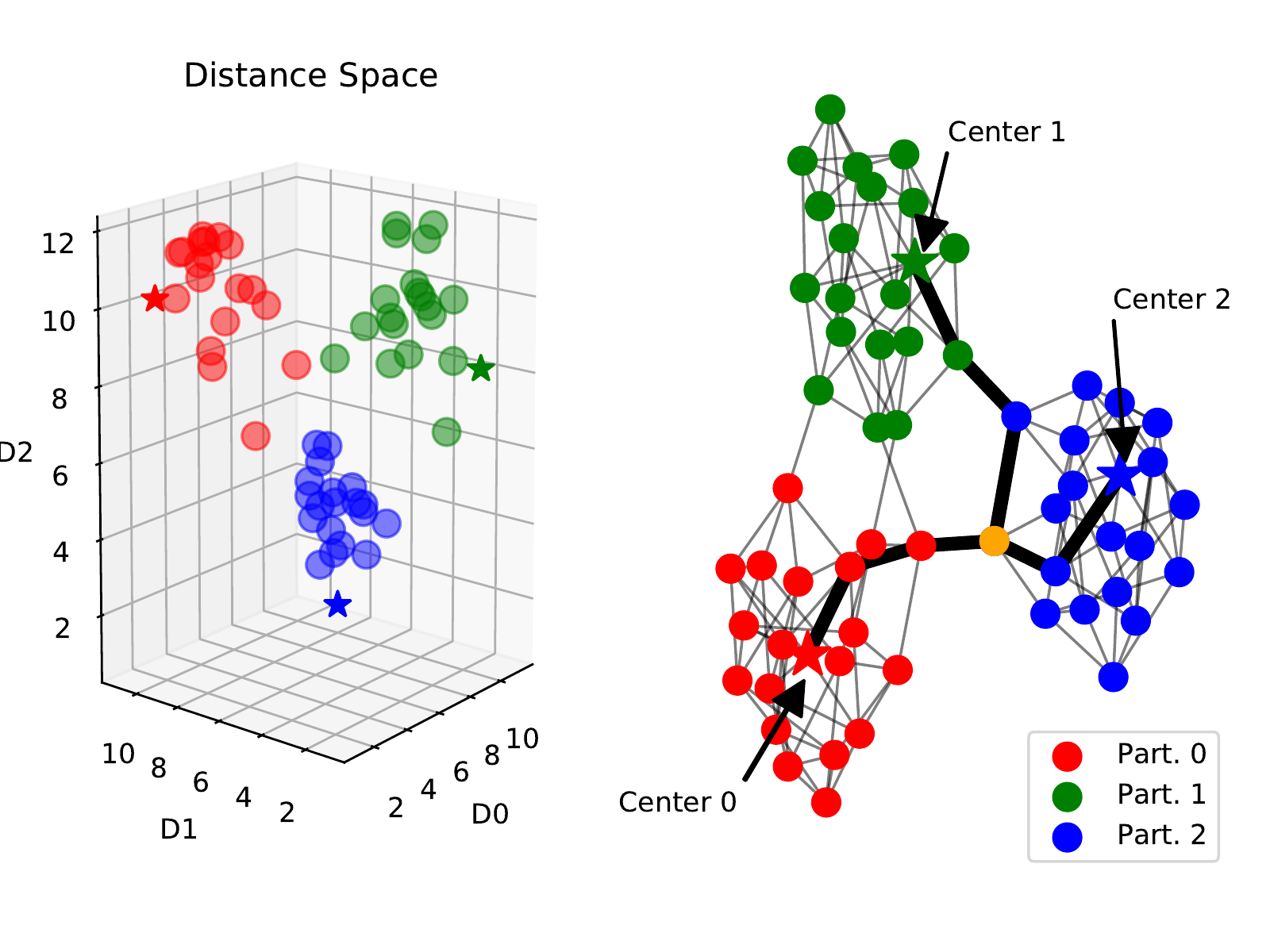}
    \caption{An illustration of the method. (a) A 3-dimensional space representing community distances is created. Each axis corresponds to the shortest path distance from a node and the respective community seed. The starts represent the community centers. (b) In the network visualization, thick lines indicate the shortest path of the orange node to the community centers.}
    \label{fig:didact}
	\end{center}
\end{figure}

The methodology is illustrated in Fig.~\ref{fig:didact} for the case of three communities. The plot on the left shows a 3-dimensional space where the coordinates indicate the node distance to the community reference nodes. Nodes represented by the stars are the seeds used for community identification. The figure on the right shows the shortest paths from a reference node, in orange, to each seed. The paths are shown as thick lines. In this case the orange node will be classified as belonging to the blue community.

\section{Results and discussion}

We start by applying the methodology to the Zachary's karate club. Then, the methodology is validated using a artificial network model. Next, the method is used for identifying conurbations among cities.

\subsection{Zachary's karate club}

\begin{figure}
	\begin{center}
    \includegraphics[scale=0.55]{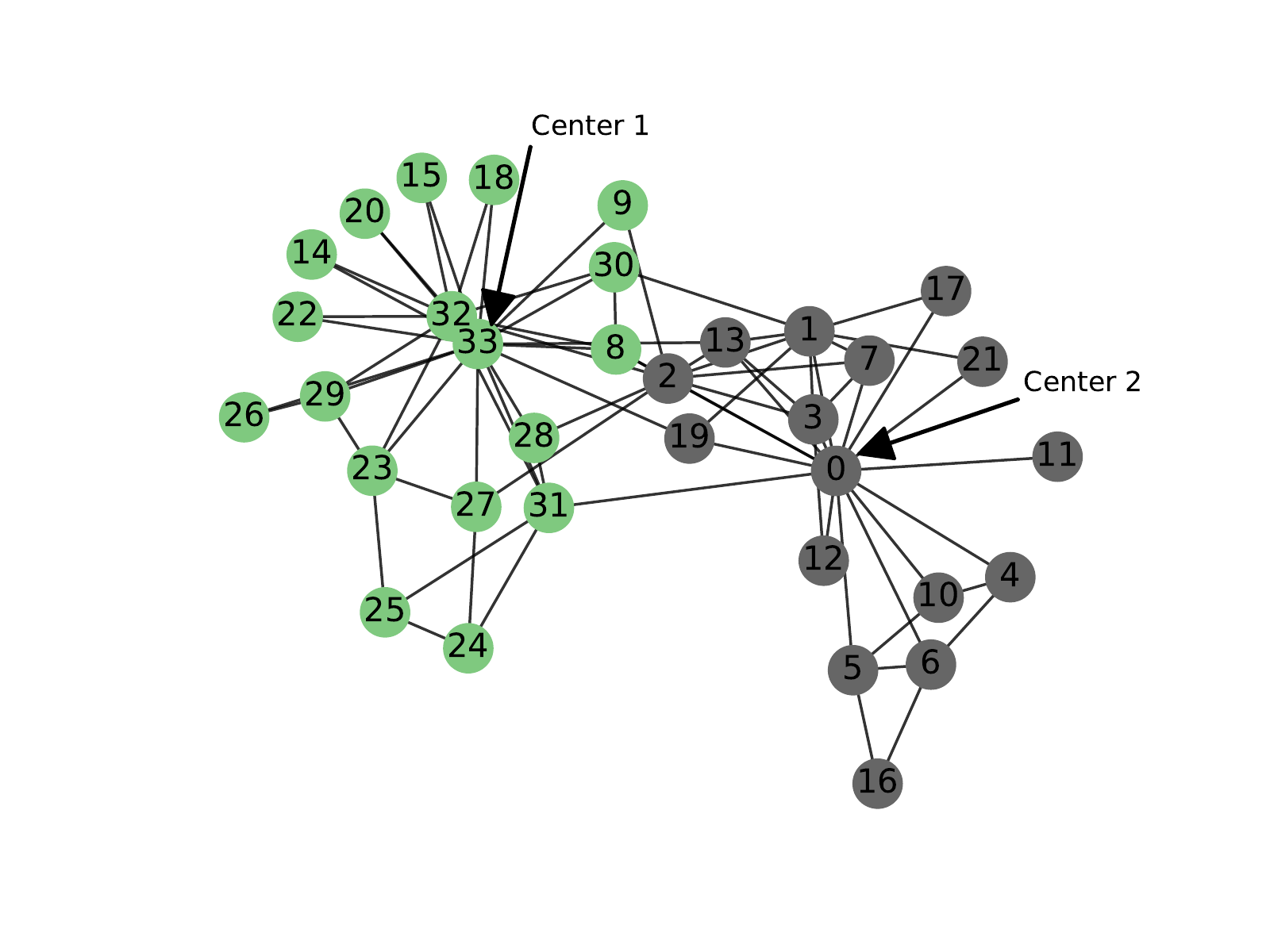}
    \caption{Zachary's karate club network partitioned correctly by our method by choosing the two highest degree nodes} 
    \label{fig:kara}
	\end{center}
\end{figure}

As a first test, the proposed method was applied to the Zachary's karate club network~\cite{zachary1977information}. The two nodes with the largest degree, $v_{i=0}$ and $v_{i=33}$, were set as reference for each community.
The topological distance was used. The detected communities are shown in Fig.~\ref{fig:kara} and coincides with the two main well-known communities of this network.

\subsection{Geographic networks}

Given that the proposed method takes into account the distances between each node and the communities seeds, it is natural to verify how the method behaves in geographic networks.

\begin{figure*}
	\centering
  	\subfigure[]{
    \includegraphics[scale=0.25]{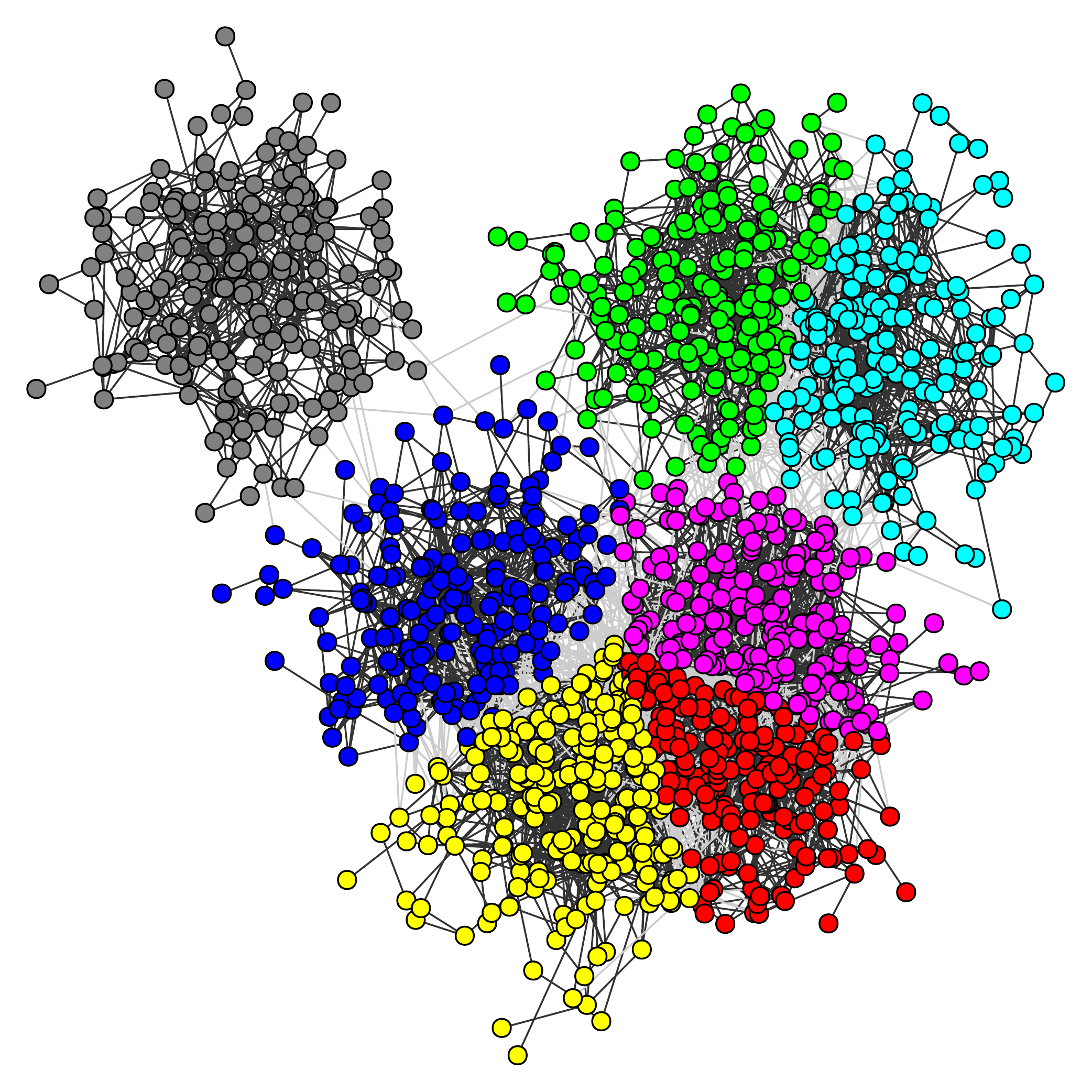} 
    \label{fig:fortuna}
  	} 
    \quad
  	\subfigure[]{
    \includegraphics[scale=0.5]{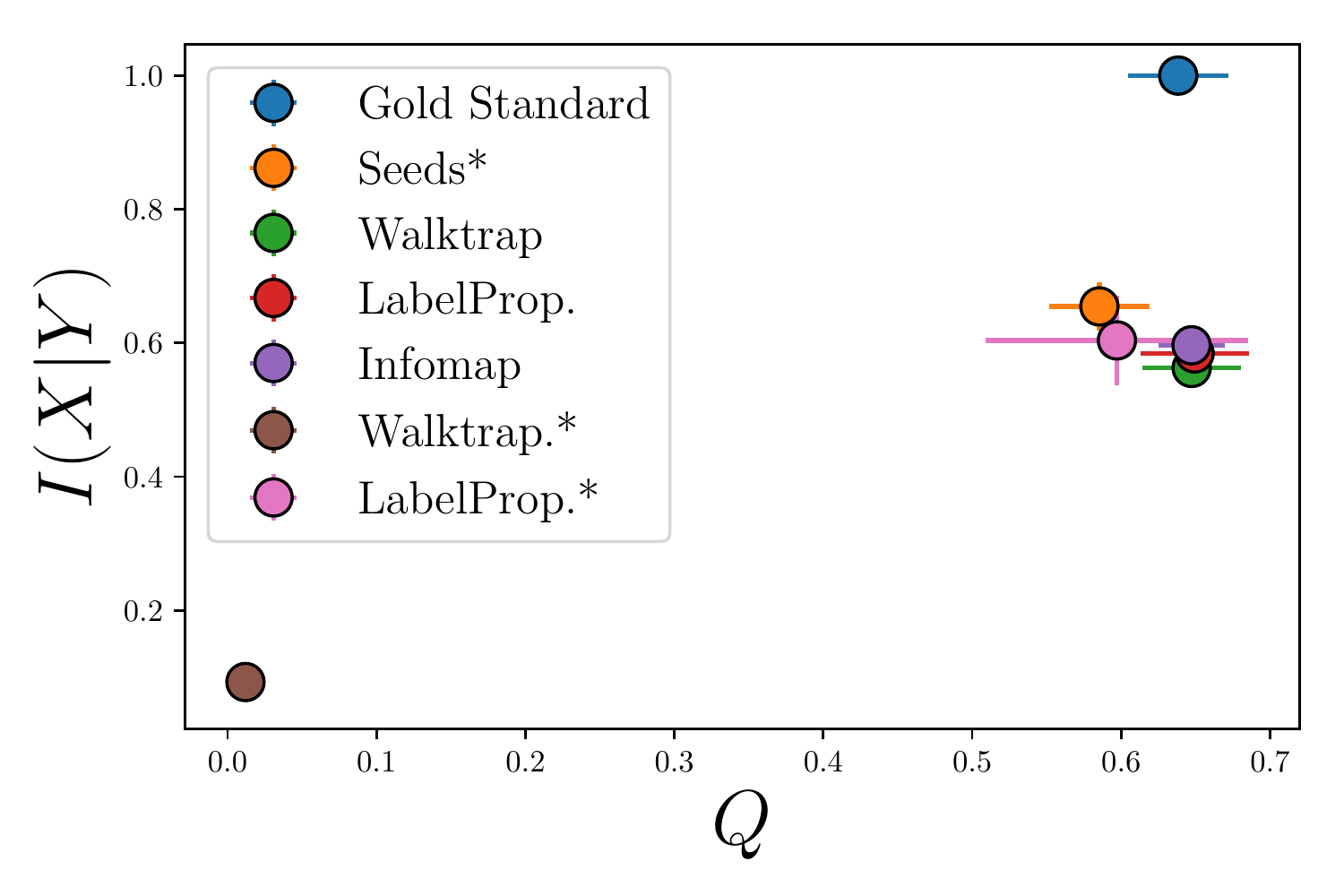}
	\label{fig:fornt_var_mix}  	
  	} 
   \caption{(a) A spatial network where the colors represents the gold-standard partitions. (b) Characterization of a community detection method on normalized mutual information between the gold-standard and modularity. The points colors represents the method used for the network partition. \emph{Walktrap*} and \emph{LabelProp*} used \emph{a priori} information about the network. }
   \label{fig:wax_gauss}
\end{figure*}

The geographical network benchmark model is generated by the following process. First we placed seed nodes according to a Poisson point process~\cite{chiu2013stochastic} and each seed was associated to a bi-dimensional gaussian distribution in order to create a probability density. Then we generated points using the probability density and connected the points using the Waxman algorithm~\cite{waxman1988routing}. The algorithm connects geographically distributed nodes based on the Euclidean distance between them. The connection probability is given as:
\begin{equation}\label{eq:wax_eq}
	p_{ij} = e^{-\beta d_{ij}}
\end{equation}
where $\beta$ is a parameter that controls the network average degree, and $d_{ij}$ is the distance between node $v_i$ and node $v_j$. The gold-standard is obtained by the Voronoi tessellation~\cite{ferenc2007sizeVoronoi}, where each cell, corresponding to a partition, is centered at the respective Gaussian mean.

Fig.~\ref{fig:wax_gauss} shows a comparison between our method and the traditional methods considered in this work, where the x-axis represents the modularity and the y-axis the normalized mutual information. The partitioning methods, Label Propagation and Walktrap, were first applied without any information about the partitions. Then, the number of communities and specific seeds (the same use in our method) were supplied to, respectively, the Walktrap and Label Propagation methods.

The partitioning of the geographical networks shows that our method, the Walktrap and the Label Propagation have similar scores in both experiments. This suggest that the Label Propagation and the Walktrap when using prior information does not work well in this kind of network. The proposed method is the closest point in this plane to the gold-standard partitions, as it reached the largest score of normalized mutual information and have a high value of modularity. In addition, the suggested method yielded small standard deviations in both axes.

\subsection{Cities networks}

\begin{figure}
	\centering
  	\subfigure[]{
    \includegraphics[scale=0.10]{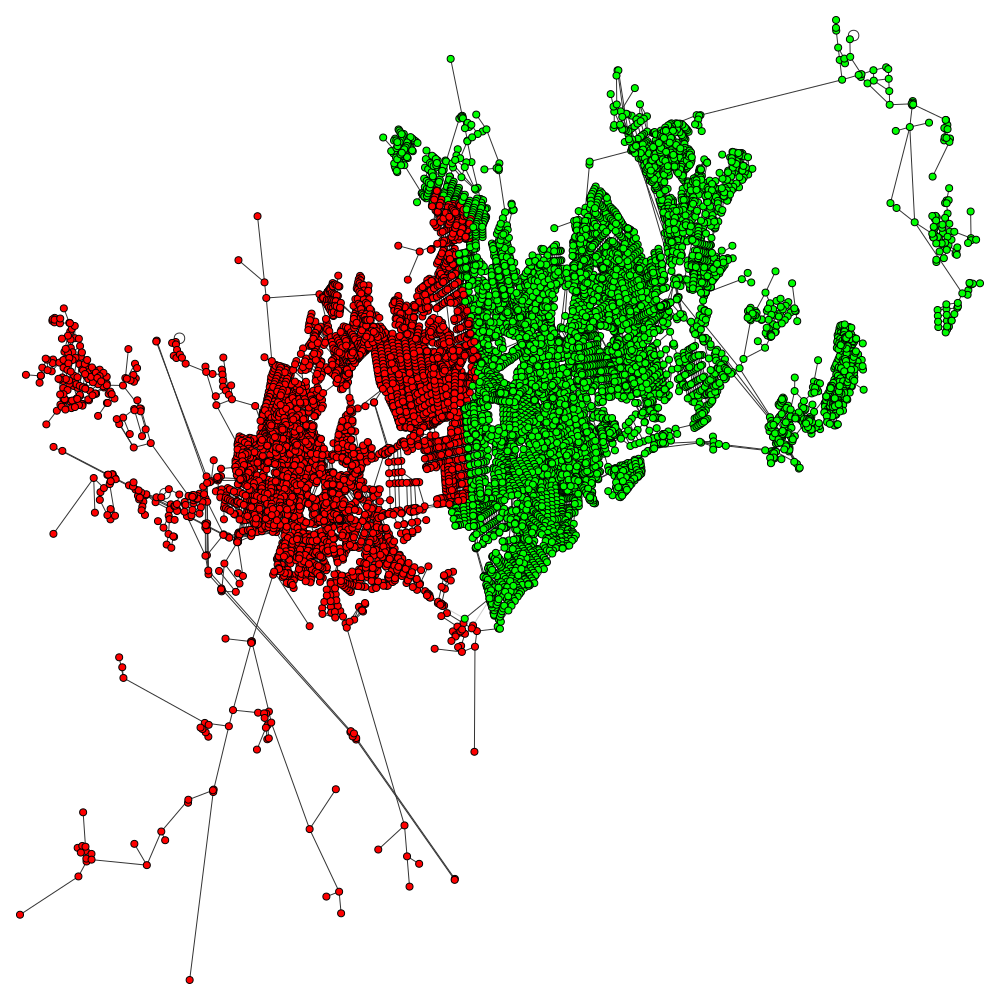} 
    \label{fig:adm_citie}
  	} 
    \quad
  	\subfigure[]{
    \includegraphics[scale=0.10]{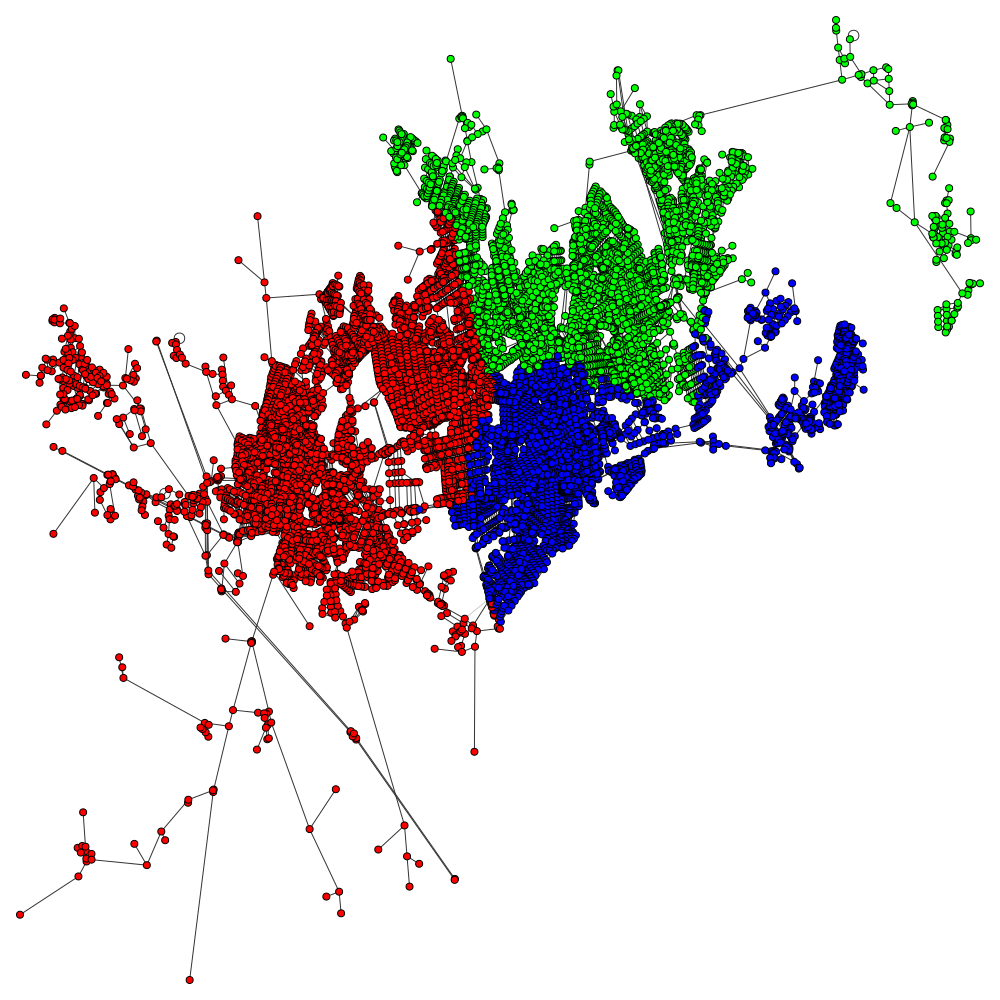} 
	\label{fig:seed}  	
  	} 
  	\caption{Cities results.}
	\label{fig:cities}
\end{figure}

A possible application for our community detection method concerns to the identification of regions in cities networks.
In this study we apply our method to identify two cities in a conurbation.
We obtained the cities networks using the python library osmnx~\cite{boeing2017osmnx}.
We choose the cities of Americana and Santa Bárbara D'Oeste in the state of São Paulo, Brazil.
The seeds were selected considering the proximity to the cities administrative centers.
Fig.~\ref{fig:adm_citie} shows the two administrative areas of the cities: in green the city of Americana, and in red the city of Santa Bárbara D'Oeste.
Fig.~\ref{fig:seed} shows the segmentation of the conurbation in two regions, note that two centers had to be used in order to
achieve a good adherence between the obtained community and the original administrative region.  This suggests that the administrative labeling is not completely compatible with the urban topology of the intersecting cities.
The suggested method has good application potential for characterizing the inner structure of cities.
For example, we can use as seeds hospitals, subways or supermarkets. Using this kind of seed, our method will partition the city network where the generated communities will have a straightforward conceptual interpretation. 



%

\section{Conclusion}

Community detection in complex networks remains an interesting research subject. 

In this paper we introduced a new approach to network partition that is specified in terms of conceptually intuitive vectors of distance to the centers of each community, establishing a kind of community-based coordinate system.  The centers are nodes that have an important role inside of the community structure, being chosen using prior information about the network. 

In the geographical benchmark partition, the proposed method achieved the highest value of mutual information allowing the correct identification of the regions.

The obtained results suggest that the method proposed in this work can be used in several kinds of network partition problems, highlighting situations that have geographic features. Furthermore, a community-based coordinate system is obtained as a byproduct, that can provide additional information about the partitions and the network structure.

Several future developments could be considered, like using other distance metrics and exploring the information provided by the distance coordinate system.

\acknowledgments
C. H. Comin thanks FAPESP (grant no. 18/09125-4) for financial support. L. da F. Costa thanks CNPq (grant no. 307333/2013-2) and NAP-PRP-USP for sponsorship. This work has been supported also by FAPESP grants 11/50761-2 and 2015/22308-2.

\bibliographystyle{apsrev}
\bibliography{references1.bib}
\end{document}